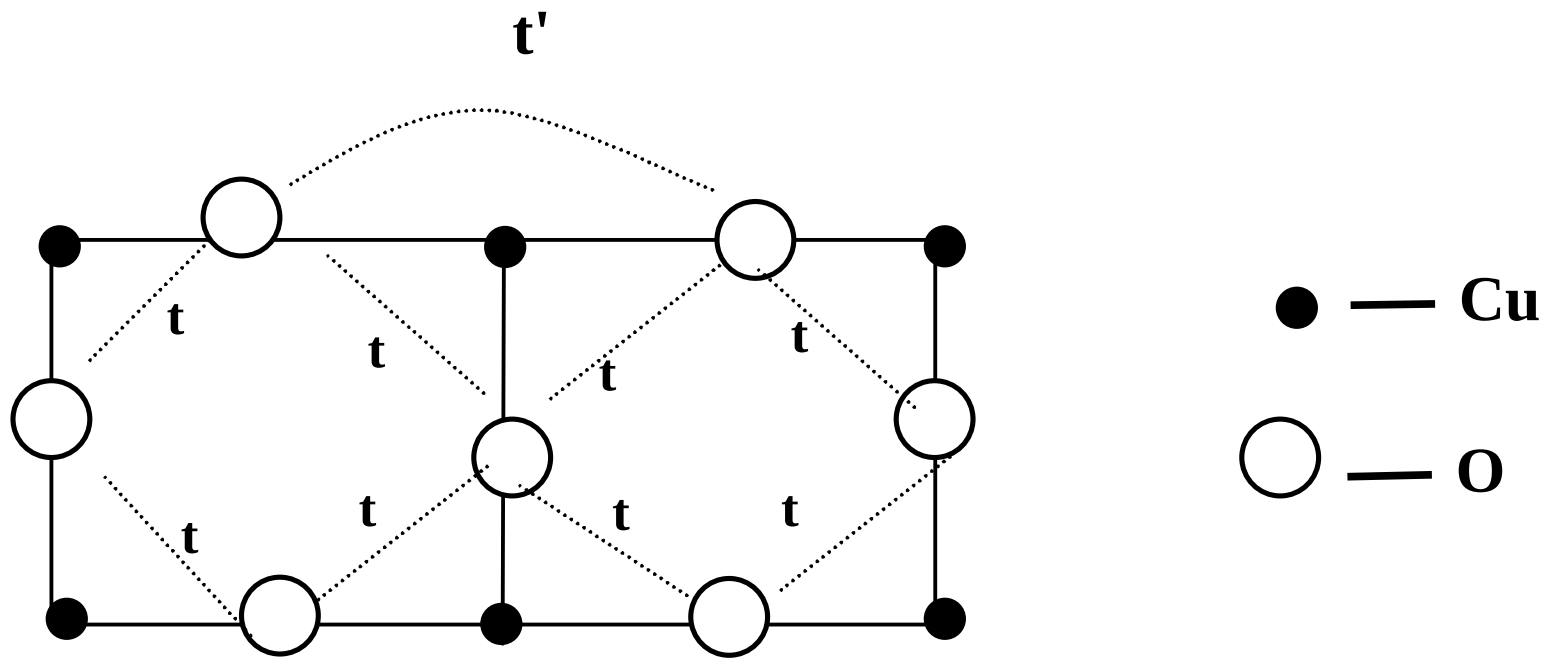

**Fig.1**

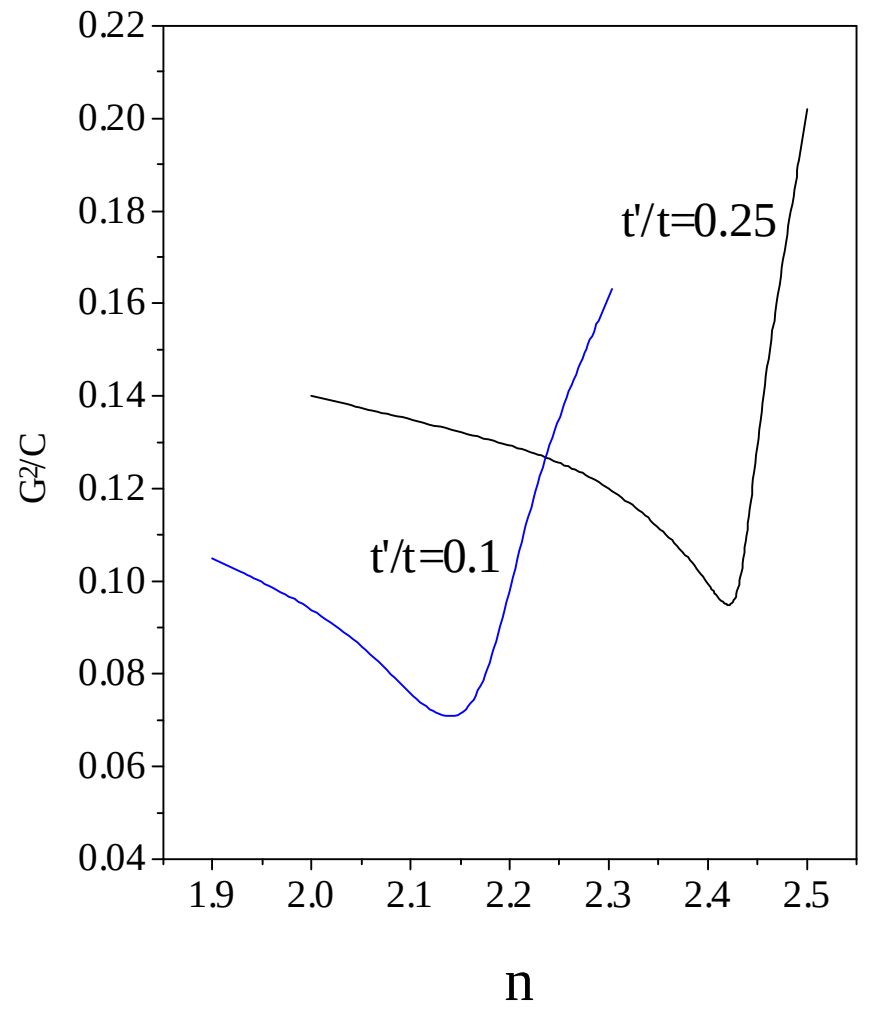

**Fig. 2**

# Pseudogap and its influence on normal and superconducting states of cuprates

I. Chaudhuri[1,*], A. Taraphder[1,2] and S. K. Ghatak[1]
[1]Department of Physics and Meteorology, [1]Center for Theoretical Studies, I.I.T, Kharagpur, 721302

## Abstract

A model incorporating simultaneous superconducting and lattice instabilities has been studied in detail to estimate the nature of coupling and inter-play between them. The phase diagram is obtained in the temperature-filling plane at different values of the parameters of the model. It is found that a pseudogap develops in the distorted phase that inhibits the appearance of the superconducting transition. The superconducting instability is strongest for the regime of filling where the van Hove singularity in the 2D density of states is close to the chemical potential. The pseudogap, developed in the distorted phase, is a function of temperature via the temperature dependence of the distortion itself. Transport properties, namely resistivity and thermopower are found to be strongly dependent on the variations of the pseudogap.

## 1. Introduction:

The existence of a pseudogap in the electronic spectra of high-$T_C$ superconductor below a certain temperature is considered to be one of the most important features in the cuprates [1]. A dip-like structure in the density of states defines the pseudogap. Many experiments, like nuclear magnetic resonance (NMR) [2], angle resolved photoemission spectroscopy (ARPES) [3], specific heat [4], electron-tunneling spectroscopy (ETS) [5], scanning tunneling spectroscopy (STM) [6] have provided ample evidence for a gap-like structure in the electronic excitation spectrum. All the high-$T_C$ cuprates have Cu-O plane as their common building block. In the absence of doping the $CuO_2$ plane has strong antiferromagnetic correlation and the insulating antiferromagnetic state is converted into a metallic paramagnetic one with doping. It is believed that the antiferromagnetic correlation still persists in metallic state. The nearly 2D structure of Cu-O layer produces sharp van Hove - like structure in the density of states (DOS). The structural transition from a tetragonal to an orthorhombic phase and the associated lowering of the symmetry has long been considered important in the high $T_C$ superconductors. The origin of the pseudogap in cuprates is still not well established although number of possible mechanisms have been considered based on any one of above features. It is assumed that residual antiferromagnetic correlation in doped system leads to a gap in the DOS, which is generally referred to as spin-gap [7]. Another model views the pseudo-gap as consequence of 'precursor pairing ' where the formation of pairs occurs first and the phsae coherence is established at $T_c$ [8]. An alternative to this view is 'preformed pair' at higher temperature and below the critical temperature they undergo Bose –Einstein condensation [9]. The resonance scattering by superconducting fluctuation is thought of as a possible origin for the pseudogap [10]. The existence of the charge density-wave initiated by van Hove – like singularity in the density of states is also considered as the source of this gap [11]. Assuming a simple model for gap in the excitation spectrum in normal state an interplay between gap and superconductivity has been examined by Nozieres and Pistolesi [12]. This model has recently been extended by considering the momentum dependence of the gap and the phase diagram for superconducting and pseudogap phase has been derived assuming a phenomenological T-dependence of the pseudogap [13]. Yet another possible origin of the pseudogap has been put forward based on the Jahn-Teller - like distortion [14]. The importance of structural distortion -- a characteristic feature of oxides (very strong distortion is known to occur in maganites) is borne out by the experimental facts [15]. The susceptibility to distortion depends on the carrier concentration. In underdoped systems distortion appears at higher temperature and the magnitude of distortion is large. The transition temperature for spontaneous distortion $T_p$ decreases and appears to meet the superconducting transition temperature near optimal doping [15]. For overdoped compounds the distortion seems to vanish. The presence of distortion in these systems is found to reduce both transition temperature $T_c$ and the superconducting order parameter [16]. When superconductivity sets in the growth of the distortion is either arrested or suppressed. The discontinuous change in thermal expansion observed in cuprates [17] can be related to the suppression of growth of distortion. In this article we present in detail results on pseudogap, transport properties and the phase diagram that follow from the proposed model [14].

## 2. Model and calculations:

We consider that the doped charge carriers reside in the $CuO_2$ layer. The electronic states of a layer are normally described by a model with three states (two bonding states $p_x$ and $p_y$ of O and one d-state of Cu) [18]. However, it is also believed that the charge carriers are predominantly in the p-state of O. So we assume a description of the $CuO_2$ layer in terms of only the two oxygen p-states. In the high temperature phase the two states ($p_x$, $p_y$) are degenerate due to the symmetrical position of O. Any asymmetry will create different crystal potential which in turn remove the degeneracy of the state in unit cell. The distortion within the unit cell can lift the degeneracy. The existence of the spontaneous distortion depends on the competition between the lowering of the electronic energy and the increase in elastic energy. Superconductivity will be affected in presence of distortion. The model Hamiltonian to describe the existence of distortion and its interplay with superconductivity within this two-state model can be written as [14]

$$H = H_0 + H_P$$

where $H_o$ is the effective one-electron Hamiltonian.

$$H_0=\sum_{\alpha\sigma}\varepsilon_{p\alpha}\, p^{+}_{i\alpha\sigma}\, p_{i\alpha\sigma} + \sum_{ij} t_{ij}\left(p^{+}_{i\alpha\sigma}\; p_{j\alpha'\sigma} + h.c\right) + \sum_{ij} t'_{ij}\, p^{+}_{i\alpha\sigma}\; p_{j\alpha\sigma} \qquad (1)$$

Here $\alpha$ represents the orbital (x,y) and i and j denote the sites. The site energy $\varepsilon_{p\alpha}$ is different for different orbitals the in presence of distortion in Cu-O cell (Fig-1). The site energy of the orbital will increase (decrease) due to elongation (compression) of the axis and can be approximated as $\varepsilon_{p\alpha} = \varepsilon_{p\alpha}^{o} \pm(\partial\varepsilon_{p\alpha}/\partial\ e)\ e \approx \varepsilon_{p\alpha}^{o} \pm G\ e$ ,G being the electron–lattice interaction strength. The nearest neighbour hopping integral between $p_x$ and $p_y$ oxygen orbital is represented by t, and t' is the next nearest neighbour hopping mediated through d-orbital of the Cu atom. The Hamiltonian

$$H_p = - \sum_{\{k\},\{\alpha\}} g^{\alpha\alpha',\alpha_1\alpha_2}_{kk'}\;\; p^{+}_{k\uparrow\alpha}\, p^{+}_{-k\downarrow\alpha'}\, p_{-k'\downarrow\alpha_1}\, p_{k'\uparrow\alpha_2} \qquad (2)$$

represents retarded pairing interaction of the BCS type between the same as well as different p-orbitals with pairing potentials given by $g^{\alpha\alpha',\beta\beta'}_{kk'}$.

It is assumed that the pairing strength is finite over a cut-off energy $\Omega$ about the Fermi level. Using the BCS mean-field approximation for $H_p$ the total Hamiltonian reduces to

$$H = \sum_{k,\alpha}\varepsilon_{p\alpha}\, p^{+}_{k\alpha\sigma}\, p_{k\alpha\sigma} + \sum_{k,\alpha}\varepsilon_{k}\, p^{+}_{k\alpha\sigma}\, p_{k\alpha'\sigma}$$
$$+ \sum_{k,\alpha_1\alpha_2}\left[\Delta^{\alpha_1\alpha_2}\, p^{+}_{k\uparrow\alpha_1}\, p^{+}_{-k\downarrow\alpha_2} + \Delta^{\alpha_1\alpha_2}\, p_{-k\downarrow\alpha_2}\, p_{k\uparrow\alpha_1}\right] \qquad (3)$$

where

$$\varepsilon_k = 4t\sin k_x a/2\,\sin k_y a/2$$
$$\varepsilon_{p\alpha} = 2t'\left(\cos k_x a + \cos k_y a\right) - \mu \pm Ge, \quad \text{(+) for } \alpha=1, \text{(-) for } \alpha=2 \qquad (4)$$

and

$$\Delta^{\alpha_1\alpha_2} \equiv \Delta^{\alpha_1\alpha_2}_{k'} \equiv \sum_{k,\alpha,\alpha'} g^{\alpha\alpha',\alpha_1\alpha_2}_{kk'}\left\langle p^{+}_{k\uparrow\alpha}\, p^{+}_{-k\downarrow\alpha'}\right\rangle$$

The coupled Hamiltonian is quadratic in electron operator and is solved in the broken symmetry state using standard methods leading to the set of four equations for the Green functions $G_{l,m} = -\langle\langle p_{k\sigma l}, p^{+}_{k\sigma m}\rangle\rangle$ and $f^{+}_{lm} = -\langle\langle p^{+}_{-k\downarrow l}\, p^{+}_{k\uparrow m}\rangle\rangle$

$$\begin{pmatrix} \omega-\varepsilon_{p1} & -\varepsilon_k & \Delta^{11} & \Delta^{12} \\ -\varepsilon_k & \omega-\varepsilon_{p2} & \Delta^{12} & \Delta^{22} \\ \Delta^{11} & \Delta^{12} & \omega+\varepsilon_{p1} & \varepsilon_k \\ \Delta^{12} & \Delta^{22} & \varepsilon_k & \omega+\varepsilon_{p2} \end{pmatrix} \begin{pmatrix} G_{11} \\ G_{12} \\ f_{11} \\ f_{12} \end{pmatrix} = \begin{pmatrix} 1 \\ 0 \\ 0 \\ 0 \end{pmatrix} \tag{5}$$

Here

$$G_{11} = \sum_{k,i} \frac{A_i}{\omega-\omega_i} \qquad (i=1,2,3,4) \tag{6}$$

Where $A_i$, are the spectral weights

$$A_i = (\omega_i - \varepsilon_{p2})\left((\omega_i + \varepsilon_{p1})(\omega_i + \varepsilon_{p2}) - \varepsilon_k^2\right) - \Delta^2(\omega_i + \varepsilon_{p1}) - \Delta_3^2(\omega_i + \varepsilon_{p2}) + 2\Delta\Delta_3\varepsilon_k \tag{7}$$

The corresponding eigenvalues, $\omega_i$ are

$$\omega_i^{\,2} = \pm\frac{1}{2}\left[\mathrm{A} \pm \omega_0\right] = \pm\omega_{\pm}^{2} \tag{8}$$

where $\omega_0 = \sqrt{\mathrm{A}^2 - 4\mathrm{B} - 4\mathrm{C}}$

with

$\mathrm{A} = \varepsilon_{p1}^2 + \varepsilon_{p2}^2 + 2\varepsilon_k^2 + 2\Delta^2 + 2\Delta_3^2$, $\quad \mathrm{B} = (\varepsilon_{p1}\varepsilon_{p2} - \varepsilon_k^2)^2 \quad$ and

$\mathrm{C} = 2\Delta_3^2\varepsilon_{p1}\varepsilon_{p2} + 2\varepsilon_k^2(\Delta^2 + \Delta_3^2) + \Delta^2(\varepsilon_{p1}^2 + \varepsilon_{p2}^2) + (\Delta^2 - \Delta_3^2)^2 - 4\Delta\Delta_3\varepsilon_k(\varepsilon_{p1} + \varepsilon_{p2})$

The coupled gap parameters are given by,

$$\Delta = \Delta^{22} = \Delta^{11} = \sum_{k,\alpha,\alpha'} g^{\alpha\alpha',11} \langle p^{+}_{k\uparrow\alpha}\, p^{+}_{-k\downarrow\alpha'}\rangle,$$

$$\Delta_3 = \Delta^{21} = \Delta^{12} = \sum_{k,\alpha\alpha'} g^{\alpha\alpha',12} \langle p^{+}_{k\uparrow\alpha}\, p^{+}_{-k\downarrow\alpha'}\rangle \tag{9}$$

Assuming the following symmetries of the interaction parameters $g^{\alpha\alpha,\alpha'\alpha'}$ on physical grounds (19)

$$g^{11,11} = g^{22,22} = g_1$$

$$g^{12,11} = g^{12,22} = g^{21,11} = g^{21,22} = g^{11,12} = g^{22,12} = g_2,$$

$$g^{12,12} = g^{21,21} = g^{21,12} = g^{12,21} = g_3$$

The equations for $\Delta$ and $\Delta_3$ take the following form

$$\Delta = -[\Delta F_1 + \Delta_3 F_2]$$
$$\Delta_3 = -[\Delta F_4 + \Delta_3 F_3] \qquad (10)$$

Where $F_1 \ldots F_4$ are given by,

$$F_1 = \frac{g_1\left[-\left(\omega_+^2-\varepsilon_{p2}^2-\varepsilon_k^2\right)-\left(\omega_+^2-\varepsilon_{p1}^2-\varepsilon_k^2\right)\right]-2g_3\left(\varepsilon_{p1}+\varepsilon_{p2}\right)\varepsilon_k}{2\omega_+\left(\omega_+^2-\omega_-^2\right)}\tanh(\beta\omega_+/2)$$
$$-\frac{g_1\left[-\left(\omega_-^2-\varepsilon_{p2}^2-\varepsilon_k^2\right)-\left(\omega_-^2-\varepsilon_{p1}^2-\varepsilon_k^2\right)\right]-2g_3\left(\varepsilon_{p1}+\varepsilon_{p2}\right)\varepsilon_k}{2\omega_-\left(\omega_+^2-\omega_-^2\right)}\tanh(\beta\omega_-/2)$$

$$F_2 = \frac{-2g_3\left(\omega_+^2-\varepsilon_{p1}\varepsilon_{p2}-\varepsilon_k^2\right)-2g_1\varepsilon_k\left(\varepsilon_{p1}+\varepsilon_{p2}\right)}{2\omega_+\left(\omega_+^2-\omega_-^2\right)}\tanh(\beta\omega_+/2)$$

$$-\frac{-2g_3\left(\omega_-^2-\varepsilon_{p1}\varepsilon_{p2}-\varepsilon_k^2\right)-2g_1\varepsilon_k\left(\varepsilon_{p1}+\varepsilon_{p2}\right)}{2\omega_-\left(\omega_+^2-\omega_-^2\right)}\tanh(\beta\omega_-/2)$$

The expressions for $F_3$ and $F_4$ are obtained replacing of $g_1$ by $g_2$ in $F_2$ and $F_1$ respectively.

The equilibrium value of strain is determined from the minimization of the free energy $F = -kT\sum_{i=1,2}\ln\left[1+e^{-\beta\omega_i}\right]+\frac{1}{2}Ce^2$ where C is the appropriate elastic constant and this leads to an equation for the strain *e*,

$$1 = \frac{G^2}{C}\sum_k\left((1+\frac{\varepsilon_p^2+\Delta_3^2}{\omega_0})\,\frac{\tanh(\beta\omega_+/2)}{\omega_+}+(1-\frac{\varepsilon_p^2+\Delta_3^2}{\omega_0})\,\frac{\tanh(\beta\omega_-/2)}{\omega_-}\right) \qquad (11)$$

The chemical potential ($\mu$) is determined by the number of electrons

$$n = \int \rho(\varepsilon)\,f(\varepsilon)\,d\varepsilon \qquad (12)$$

where $f(\varepsilon)$ is the Fermi function. Eqns. (10-12) are the coupled equations to be solved self-consistently for superconducting gap, distortion and chemical potential. The electronic density of states $\rho(\varepsilon)$ is obtained from the imaginary part of the Green functions with equilibrium values of $\Delta$ , $\Delta_3$ and e obtained from the above equation at different temperatures. A finite distortion leads to psedogap in density of states. The interplay between superconductivity and distortion is then determined by self-consistent solutions of the above coupled equations (10,11).

## 3. Result and Discussions

### 3.1 Pseudogap in the normal state

In the normal state ($\Delta=\Delta_3=0$) the equation (11) for distortion reduces to

$$1=\frac{2G^2}{C}\sum_k \frac{\tanh(\beta\omega_+/2)-\tanh(\beta\omega_-/2)}{\omega_+-\omega_-} \qquad (13)$$

with energy dispersion $\omega_i$ given by

$$\omega_i=\omega_\pm=\frac{(\varepsilon_{p1}+\varepsilon_{p2})\pm\sqrt{(\varepsilon_{p1}-\varepsilon_{p2})^2+4\varepsilon_k^2}}{2}$$

In the limit $t' >> t$ eqn. (13) becomes

$$\delta = Ge=\frac{G}{C}\sum_k\left[\tanh(\beta\varepsilon_+/2)-\tanh(\beta\varepsilon_-/2)\right] \qquad (14)$$

This is the result of band Jahn-Teller effect for two-fold degenerate in $e_g$ orbitals obtained earlier [20]. In this case orbital degeneracy is removed by distortion and can create a gap in DOS for $\delta >$ bandwidth. Here in all cases the values are normalized by 1eV. The distortion creates a gap $\delta$ in the DOS, due to the removal of orbital degeneracy of the two-fold degenerate $e_g$ band. In the other limit $t' << t$, a situation assumed to prevail in HTSC oxides, the existence of the energy gap depends crucially on $t'$. The spontaneous strain at T = 0 appears when the parameter $G^2/C$ is greater than a critical value which depends on $t'$ and carrier concentration (Fig.- 2). The critical value is obtained via eqn. (13) from the condition that the transition temperature $T_p$ goes to zero. The critical value of $G^2/C$ is least for the system with the Fermi level ($E_F$) lying at the maximum of the DOS in the undistorted phase. With smaller values of $t'$ the lower limit of the critical value goes down. For other values of concentration, when the Fermi level shifts from the peak position of DOS, the critical value increases sharply from its lowest value. This is related to the sharp variation of the DOS around its maximum. The lowering of electronic energy is a strong function of the availability of electronic states around $E_F$ in the undistorted phase. In the presence of spontaneous distortion the density of states around $E_F$ is decreased compared to that in the undistorted phase and the parameter $\delta$ is a measure of pseudogap in the DOS. Referring the system as optimally doped ($n_o$) when the $E_F$ is at the maximum of DOS the variation of $\delta$ with n around $n_o$ is depicted in Fig.3. For underdoped system with $n < n_o$, $\delta$ increases and passes through a maximum which shifts towards lower values as the parameter $G^2/C$ goes up. For overdoped case $\delta$ decreases monotonically with n. This asymmetrical relation of $\delta$ with n results from the finiteness of t' and the maximum of $\delta$ tends towards $n_o$ as the magnitude of t' goes down. The thermal variation of $\delta$ is obtained from eqns. (12) and (13). At low temperature it falls slowly and vanishes sharply at the transition temperature $T_p$ (Fig. 3 inset). The dependence of $T_p$ on carrier concentration n is shown in Fig. 4a. The arrow refers to the system for $n_o$ for t'/t = 0.1 and 0.25. The maximum value of $T_p$ occurs at $n < n_o$ for both cases. With reduction of t' the difference between $n_o$ and n for maximum $T_p$ is lowered. We note that $T_p$, like δ (T = 0), is a maximum for $n = n_o$ in the absence of t'. The transition temperature is a strong function of parameter $G^2/C$ (Fig. 4b) particularly for $n < n_o$.

The DOS $\rho(\varepsilon)$ is easily obtained from the imaginary part of the Green function. The delta functions thus appear have been replaced by Gaussians with a constant width $\tau^{-1}$ in the numerical calculation. The DOS in undistorted phase has van-Hove like structure whose position depends on the relative values of t'/t. As t′ increases from zero the position of peak shifts from zero towards band edge and returns back at higher value of t'. The pseudogap structure in the DOS for different T is displayed in Fig. 5 for t'/t = 0.25 and n = 2. The depth of pseudogap is found to be sensitive function of t', n and $G^2/C$. The pseudogap turns into a real gap at low temperature for underdoped system with a smaller value of the next neighbour hopping and/or a larger value of electron-lattice interaction. The consequences of this temperature-dependent pseudogap in the DOS on the transport properties are examined below.

**3.2 Transport Properties**

The electrical conductivity is calculated from the current-current correlation function

$$\sigma_{xx} = \int \langle\langle j_x(\tau) j_x(0) \rangle\rangle \, d\tau \tag{15}$$

where the current density operator $j_x$ is given by

$$j_x = -e\left[ \sum_{k,\alpha} \frac{\delta\varepsilon_{p\alpha}}{\delta k_x} p^{+}_{\alpha k} \; p_{\alpha k} + \sum_{k,\alpha\neq\beta} \frac{\delta\varepsilon_k}{\delta k_x} p^{+}_{\alpha k} \; p_{\alpha k} \right]$$

The first term is the contribution from nearest neighbour hopping and the second term comes from inter-orbital hopping.

The expression for the conductivity $\sigma$ can be recast in the form

$$\sigma = \int d\varepsilon \, \left(-\frac{\delta f}{\delta\varepsilon}\right) L(\varepsilon)$$

where the conductivity function *L(ε)* can be expressed in terms of the spectral functions $A_{\alpha\beta} = \mathrm{Im}\, G_{\alpha\beta}$ ,

$$L(\varepsilon) = e^2 \sum_k \left[ \left(\frac{\delta\varepsilon_{p\alpha}}{\delta k_x}\right)^2 \left(A_{11}^2 + A_{22}^2 + A_{21}^2\right) + \left(\frac{\delta\varepsilon_k}{\delta k_x}\right)^2 \left(2A_{11}A_{12} + A_{12}^2 + A_{21}^2\right) \right] \tag{16}$$

Thermoelectric power S can also be expressed in terms of L(ε)

$$S = -\frac{k}{e}\left[\frac{1}{kT}\,\frac{\int(\varepsilon-\mu)\,(-\frac{\delta f}{\delta\varepsilon})\;L(\varepsilon)\;d\varepsilon}{\int\,(-\frac{\delta f}{\delta\varepsilon})\;L(\varepsilon)\;d\varepsilon}\right] \tag{17}$$

The resistivity and thermopower have been obtained numerically assuming a constant electronic relaxation time $\tau$. Fig. 6 shows the resistivity ($\rho$) vs. temperature for different electronic concentration with $G^2/C$ = 0.18 and t'/t = 0.25. These concentrations are chosen so that the Fermi level in the undistorted phase lies close to the maximum of the DOS. At high temperature $T > T_p$, $\rho$ is metallic in nature ($d\rho/dT > 0$) and is dominated by a linear temperature dependence. Resistivity changes from metallic to semiconductor - like below $T_p$ (21) and passes through a peak as T is lowered. The peak value and its position depend on the amount of distortion that in turn depends on the carrier concentration. For low carrier concentration resistivity increases sharply below $T_p$, in tune with the sharp decrease of the DOS around the Fermi level. As distortion saturates at low temperature thermal variation of resistivity slows down. The system then behaves as a poor metal with low density of states at the Fermi level. However, for a real gap, the resistivity increases almost exponentially at low T. For $T \leq T_p$, the relative excess resistivity $[\rho(T)-\rho(T_p)]/\rho(T_p) \approx [(T_p-T)/T_p]^{0.5}$ (Fig. 6 inset). The temperature region where the resistivity is semiconductor-like shrinks as more carriers are introduced in the system.

The thermoelectric power for different concentrations and same set of parameters has been obtained. Due to the factor $(\varepsilon-\mu)$ and sharply varying Fermi-window $(\partial f/\partial\varepsilon)$ energy dependence of the conductivity function around the Fermi level plays a crucial part in determining both the sign and the magnitude of thermopower. The energy level structure of L(ε) results from the k-dependence of energy dispersion . Thermopower S passes through a broad maximum (Fig. 7). With increase in carrier concentration the maximum value of S comes down and the temperature where this maximum appears shifts towards lower values.
With the onset of distortion at $T_p$ the thermopower increases. There is a cross-over from hole-like to electron-like with increase in temperature, and the cross-over temperature depends on the doping level. At high temperature, the thermopower is negative and diffusive in nature. In the other extreme $T\rightarrow 0$, S approaches zero in a linear fashion.

### 3.3 Mutual influence of pseudogap and superconductivity

The coexistence of the Jahn-Teller like distortion or charge-stripe and superconductivity is examined by solving the coupled equations (10-12) in a self-consistent manner. Among the pairing interaction parameters g, the intra-orbital interaction $g_1$ is assumed to be larger than the g's where two different orbitals are involved. For numerical calculation we take the pairing strengths $g_1$=0.35eV, $g_2$=0.5$g_1$ and $g_3$=0.25$g_1$, the cut-off energy for pairing $\Omega$=0.1eV.
Fig.8 depicts the thermal variation of strain $\delta$ for different hole concentration (n) with a fixed value of $G^2/C$ =0.18, t'/t =0.25 and normalized by the bandwidth 4t=1eV. It is seen that the growth of $\delta$ is arrested with the onset of superconductivity at $T_{co}$ ($\delta$ =0) and the extent of lowering of $\delta$ at $T<T_{co}$ is more as $T_{co}$ increases. The maximum value of strain is reached at the SC transition temperature $T_{co}$ for all doping. Beyond a concentration that depends on the value $G^2/C$ and the pairing strengths (g's) the distortion is completely suppressed (Fig.8). Below $T_c$, SC gap and distortion compete with each other and the strain decreases, however the superconductivity co-exists with distortion within a certain region of parameter space. Within coexistence region superconducting gap parameters are also diminished due to finite pseudogap (Fig.9). The extent of reduction compared to that in undistorted phase is large for higher value of $T_p$. We note that the qualitative thermal behaviour of superconducting order parameter remain essentially same.

The phase diagram for coexistence is shown in Fig. 10 for the above set of parameters. The superconducting transition temperature $T_{co}$ increases with n and attains a maximum value at a concentration $n_o$, which can be termed as the optimal concentration. Over this region of n, $T_p$ varies slowly and meets $T_c$ near $n = n_o$. The variation of $T_p$ depends strongly on next neighbour hopping t'. In presence of distortion the superconducting transition temperature is reduced in comparison to the situation where $\delta$=0 (dashed line). This is due to the fact that the pseudogap removes a number of electronic states from the Fermi level, and this in turn reduces number of states available for pairing. As the extent of such reduction is more for higher values of $\delta$, $T_C$ ($\delta\neq0$) is much reduced for the underdoped system. We note that the distorted phase in our model has a concomitant diagonal charge-stripe order as well. The local density of charge along the diagonal direction alternates in magnitude leading to this order naturally. For the overdoped case $n > n_o$, this charge-stripe order disappears in conformity with the experimental observations [22]. The optimum doping concentration and the phase diagram are strong functions of values of the parameters. The qualitative nature of the phase diagram is in tune with experimental situation [1] and theoretically obtained [13, 23]. Within the co-existence region the density of state at low temperature has interesting structure as illustrated in Fig.11. The curve (a) and (c) correspond respectively superperconducting state with no distortion and normal state with pseudogap at low temperature. The DOS in normal state (c) has broad dip around the Fermi level. In the superconducting state with $\delta$=0 (a) a gap of magnitude $\Delta_1-\Delta_3$ appears across the Fermi level and another gap -like structure of magnitude $\Delta_1 + \Delta_3$. These features also persist within the co-existence region (b) albeit with smaller magnitude as superconducting order parameters are reduced. The peaks symmetrically located at extreme are the result of interference effect of pseudogap and superconductivity. Recent experimental results seem to suggest similar features in DOS [24].

## 4. Conclusion:

A model for the structural distortion and superconductivity in the Cuprates is proposed. It is based on the idea that the on site energy of p-level of the Oxygen atoms changes due to the increase or decrease of ligand potential in the presence of orthorhombic distortion thereby gaining a net energy through the decrease of electronic energy over the increase in elastic energy. This process removes the two-fold degeneracy of the p-level in the $CuO_2$ unit cell. The strain can create an energy gap or a pseudogap where the DOS has a vH singularity. The existence of strain also produces a stripe charge order along the diagonal direction. In the presence of a pseudogap, the resistivity and thermopower show a typical metal to semiconductor transition, which is consistent with the experimental observations [21], which suggest that this upturn of resistivity below $T_p$ arises due to the formation of stripes. In the Cuprates the charge stripe and superconducting order co-exist, though they compete with each other. Such a region of co-existence is also reproduced in our model. The extent of this co-existence region strongly depends on the electron-lattice interaction parameter and the superconducting pairing potential.
The electron correlation effects are not discussed here. Further more the influence of structural distortion on SC parameter with different pairing symmetry, such as d-wave symmetry will be an interesting study.

## 5. References

*e-mail: indira@phy.iitkgp.ernet.in

## 6. Figure Captions

Fig 1. Unit cell of $CuO_2$ with the directions of hopping t and t' indicated.

Fig 2. Critical value of electron-lattice coupling strengh ($G^2/C$) vs concentration n for different values of t'/t.

Fig 3. The strain ($\delta$) at 0K as a function of electron concentration (n) for various electron-lattice coupling strength ($G^2/C$) at t'/t=0.25. The inset shows thermal variation of strain for $G^2/C$=0.18.

Fig 4a. The structural distortion transition temperature ($T_p$) vs. filling factor for t'/t=0.1(solid) and 0.25(dashed), but same value of $G^2/C$=0.14. The straight arrow for $t^{\prime}/t$=0.1 (curl for $t^{\prime}/t$=0.25) indicate that concentration for which the corresponding $E_F$ is very close to the van-Hove singularity in the DOS.

Fig 4b. $T_p$ as a function of $G^2/C$ for various values of concentration.

Fig 5. Density of states (DOS) as a function of energy for different values of temperature, the dash line for T=0, solid line for $T/T_p$=0.5 and dotted line for $T/T_p$=1 and n=2.0 (half-filled)

Fig 6. Resistivity vs temperature for different concentration for $G^2/C$=0.18 and $t^{\prime}/t$=0.25. The change in $\ln((\rho(T)-\rho(T_p))/\rho(T_p))$ with $\ln((T_p-T)/T_p)$ in the semiconductor region of the resistivity curve is shown in the inset of this figure for different concentration.

Fig 7. Thermoelectric power vs. temperature for different concentration at $G^2/C$=0.18 and $t^{\prime}/t$=0.25.

Fig 8. Thermal variation of strain for different concentration in presence of superconductivity is shown in this figure. The strain gets completely suppressed for some concentration (n=2.48).

Fig 9. Two Solid (dashed) curves represent the variation of superconducting order parameters $\Delta_1$ and $\Delta_3$ with temperature for $\delta \neq 0$ ($\delta$=0)

Fig 10. Phase diagram for superconducting and pseudogap states. The solid (dash-dotted) curve represents the superconducting $T_c$ ($T_{CO}$) $\delta \neq 0$ ($\delta$= 0) vs carrier concentration n. The dashed curve represents the variation of $T_p$ with concentration in the normal state.

Fig 11. (b) Displays the DOS for coexistence state, shifted vertically 8.5 units, (a) DOS for superconducting state ($\delta$= 0) and (c) for normal state, moved vertically 14 units for comparison.

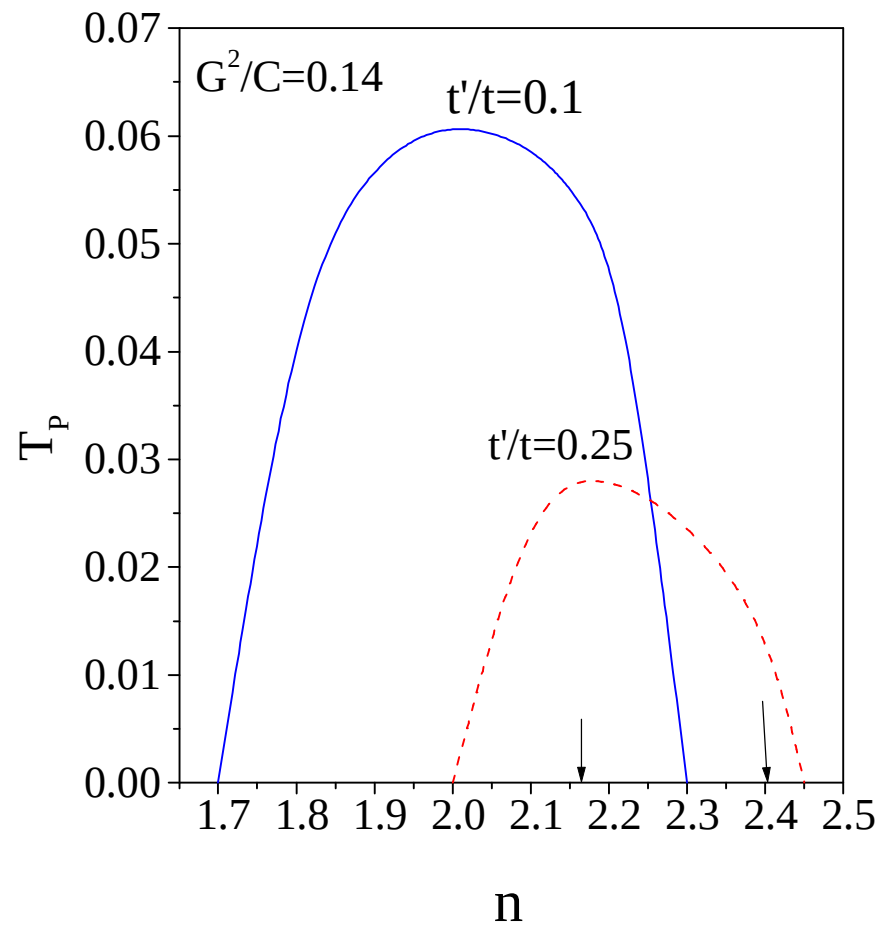
G²/C=0.14
t'/t=0.1
t'/t=0.25
0.07
0.06
0.05
0.04
0.03
0.02
0.01
0.00
T_p
1.7 1.8 1.9 2.0 2.1 2.2 2.3 2.4 2.5
n

**Fig. 4a**

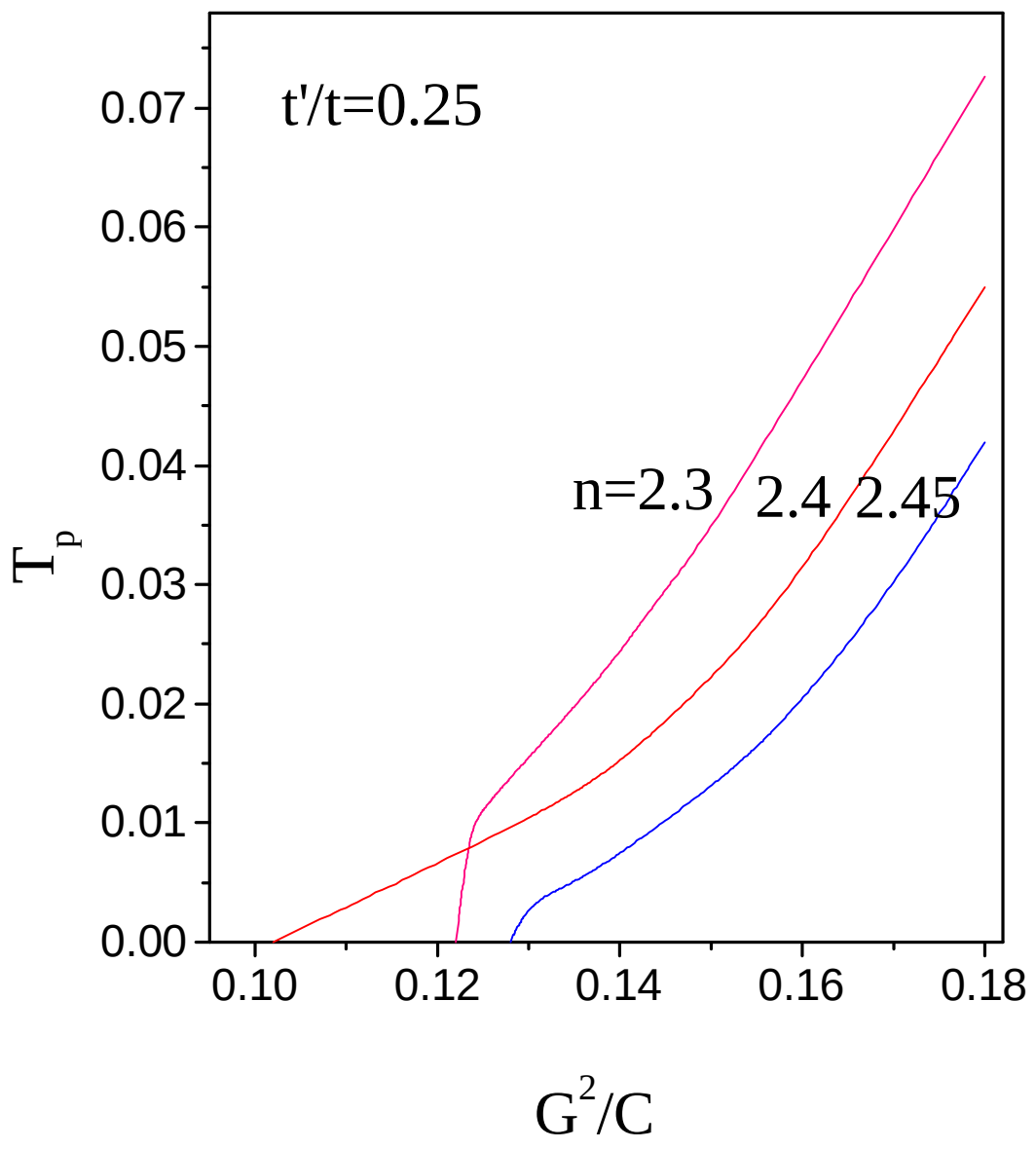

**Fig. 4b**

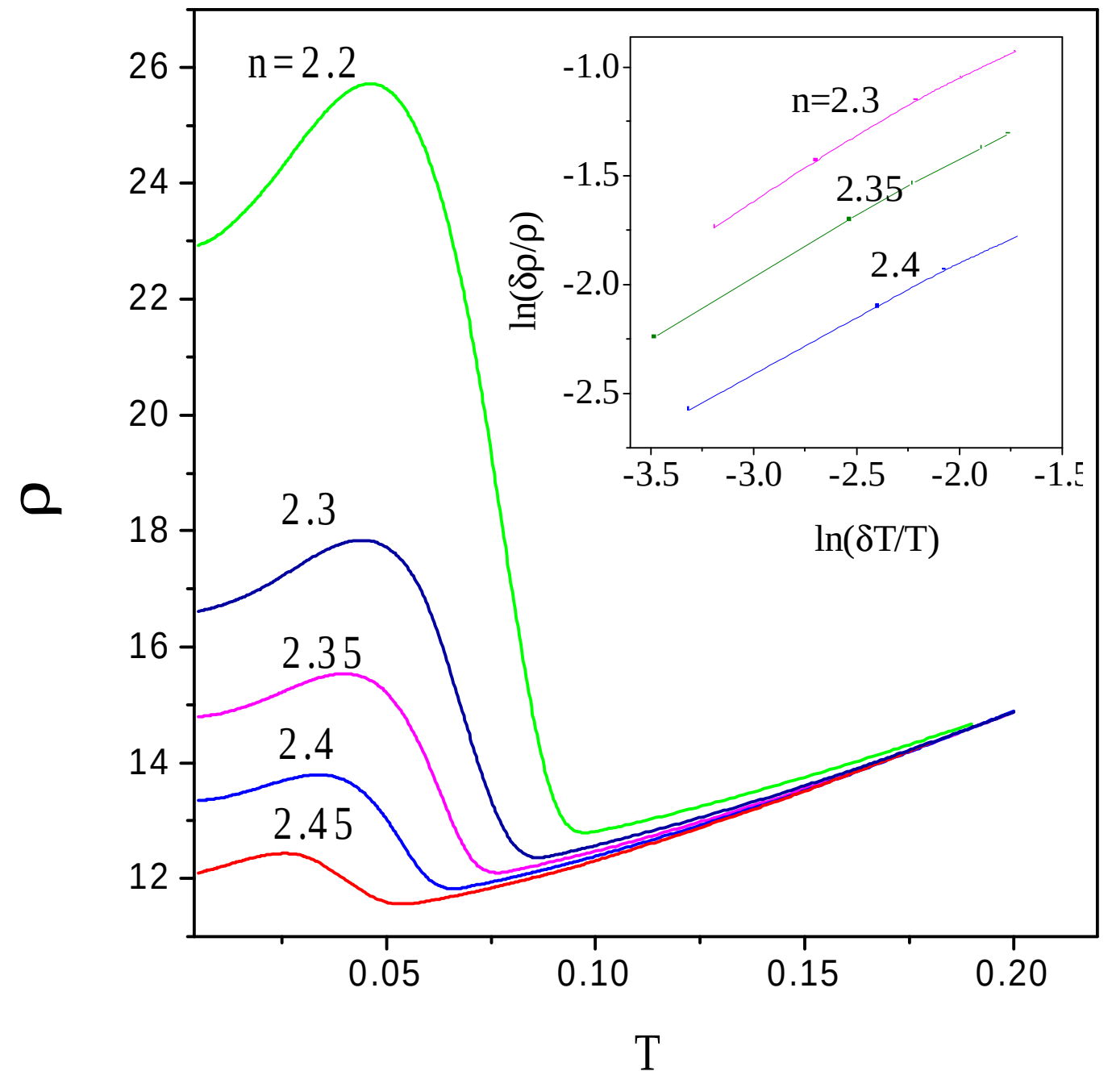

**Fig.6**

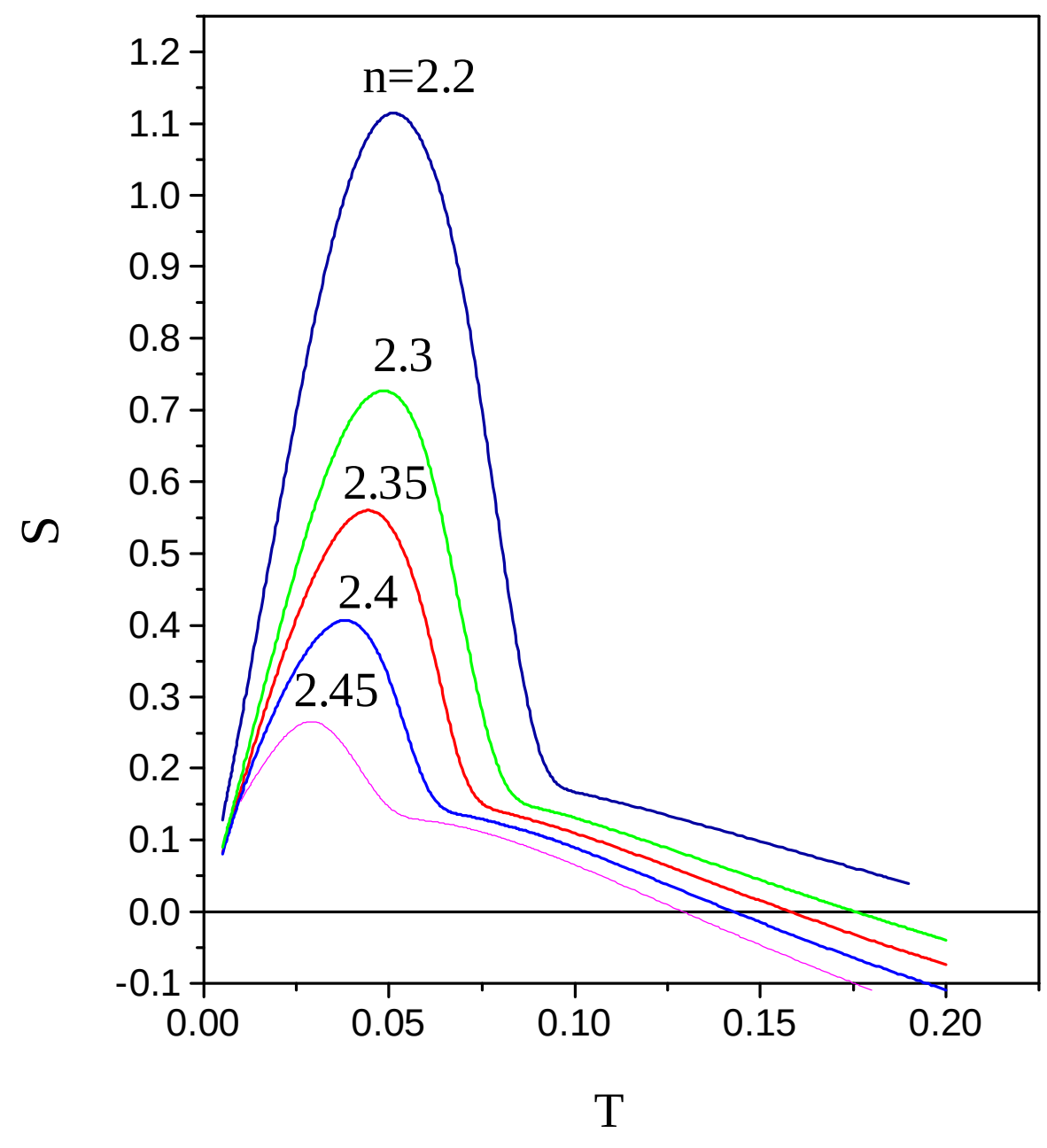

**Fig. 7**

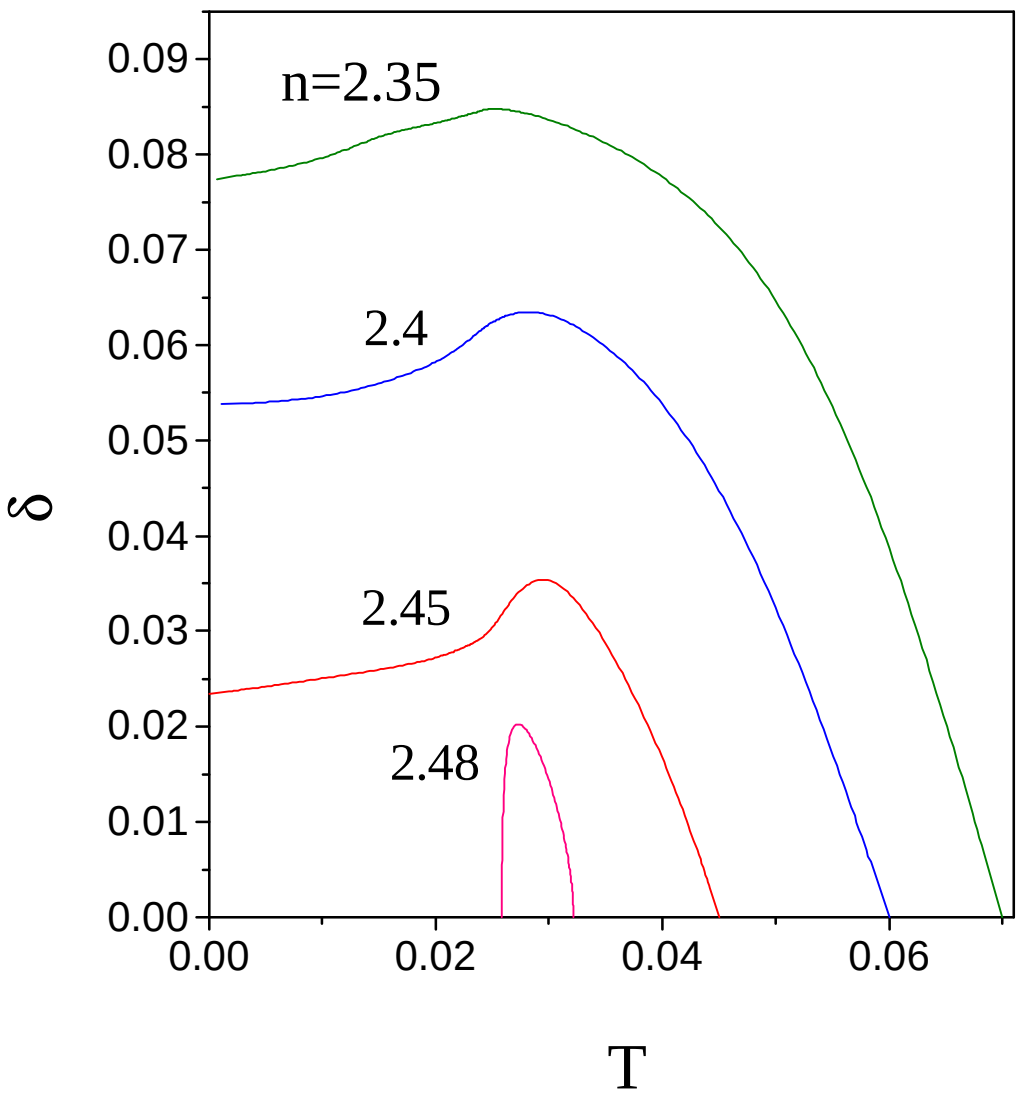

**Fig. 8**

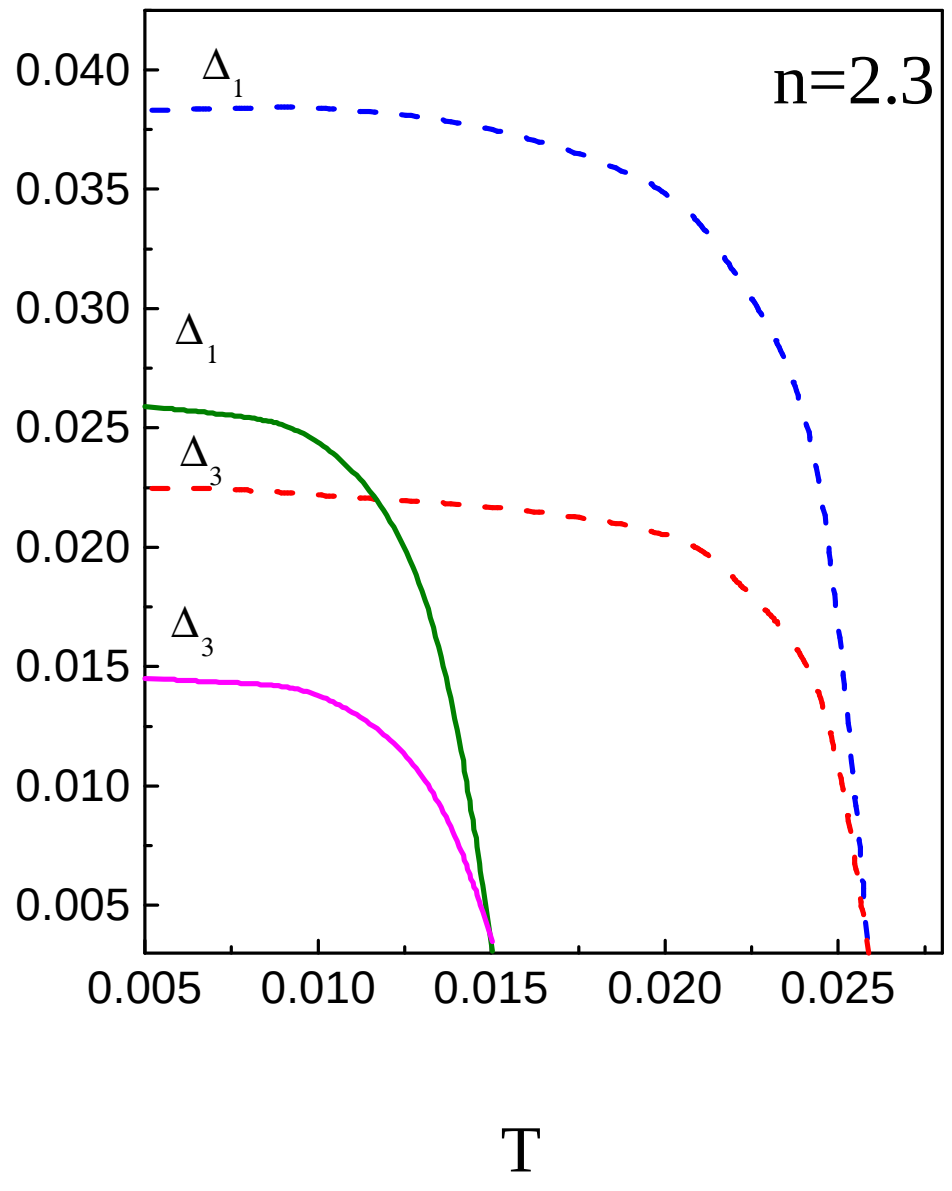

**Fig. 9**

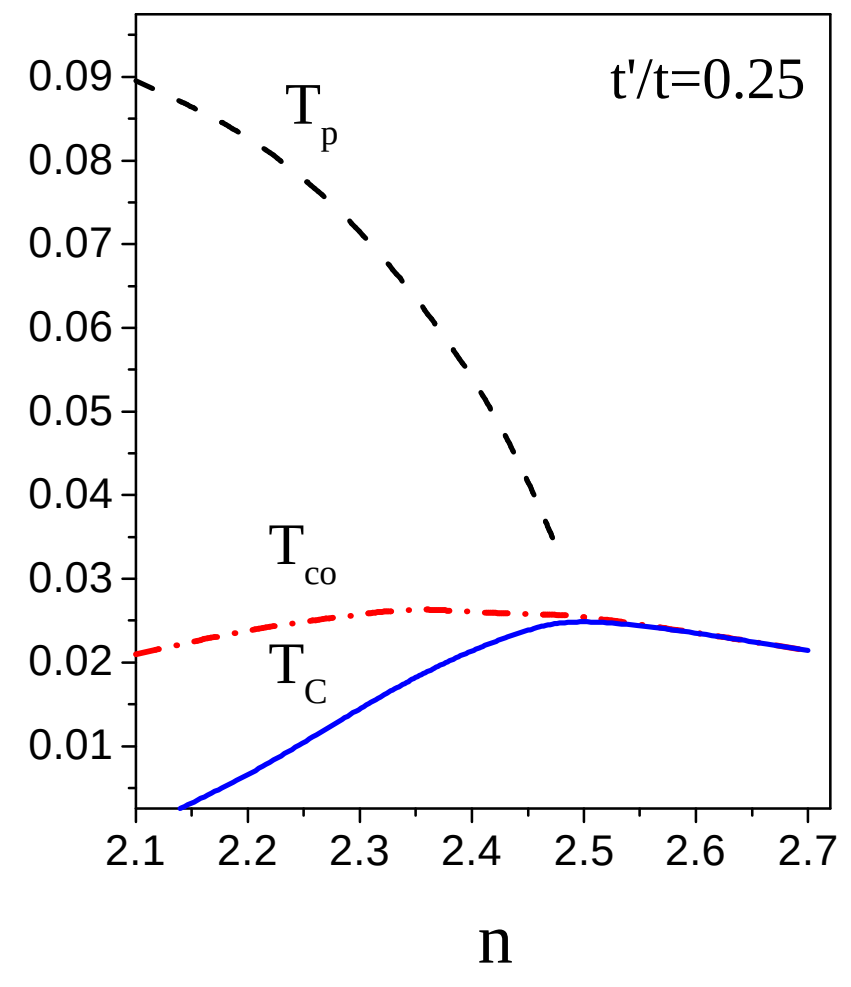

**Fig. 10**

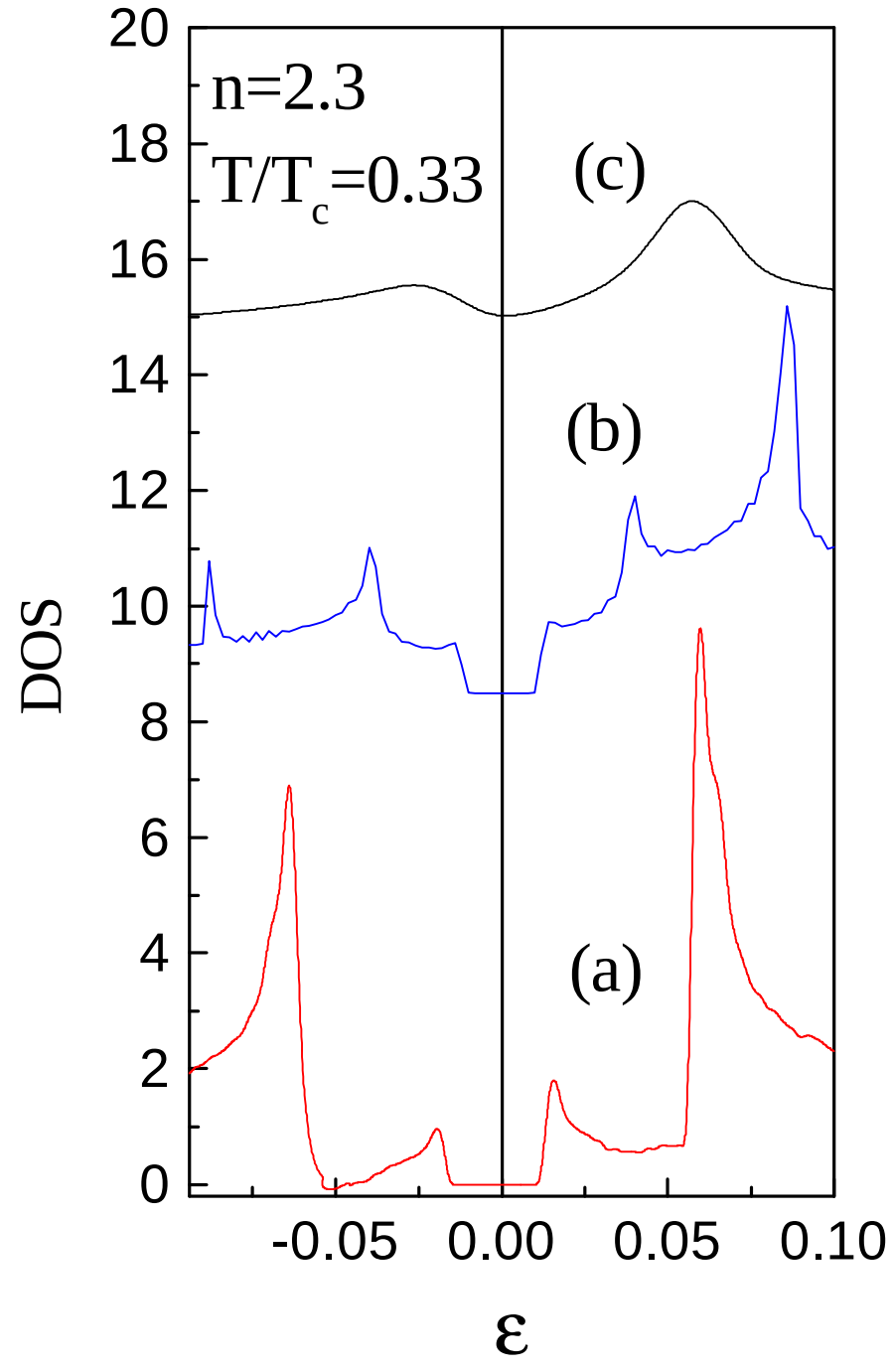

**Fig.11**

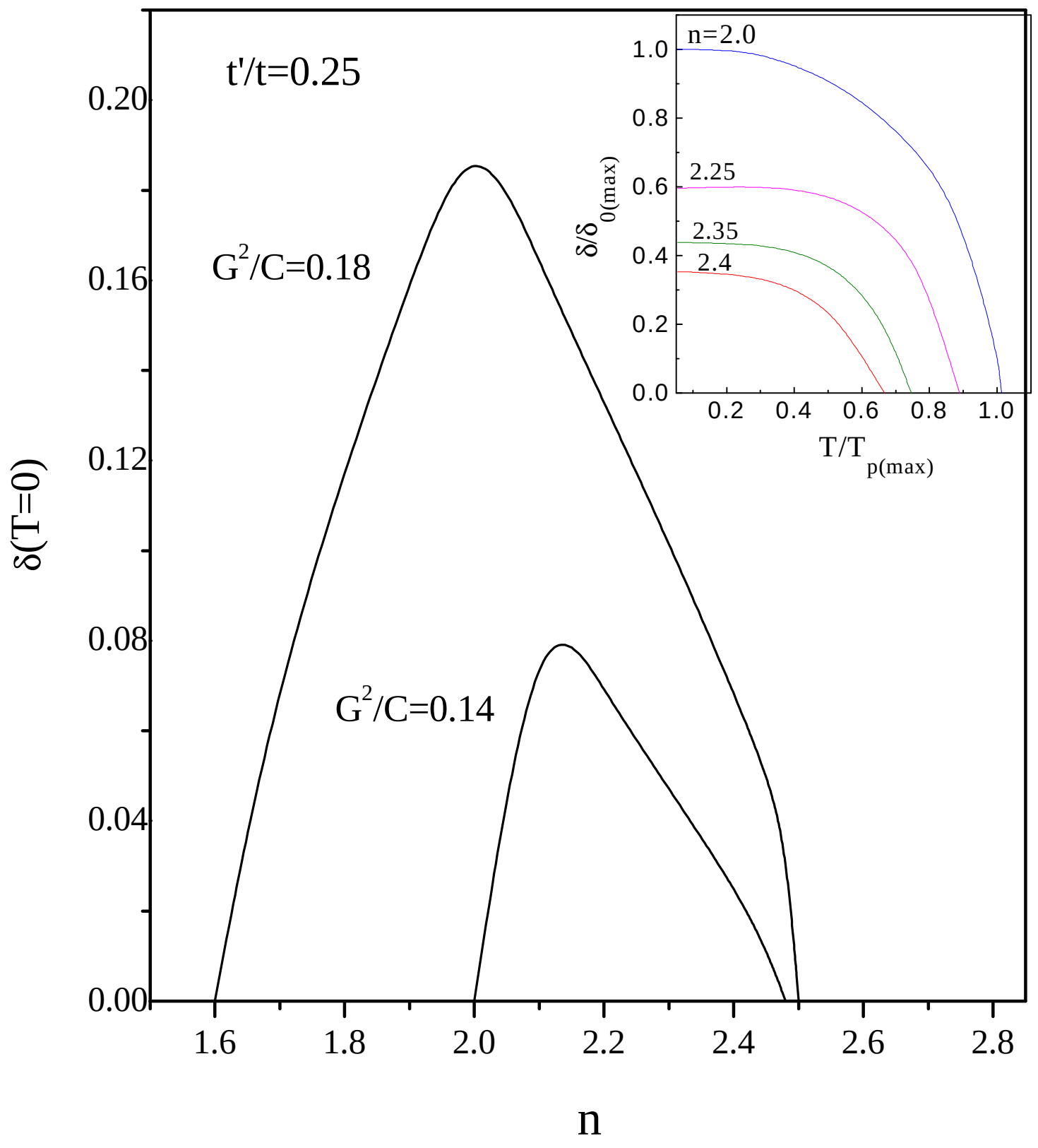

**Fig. 3**

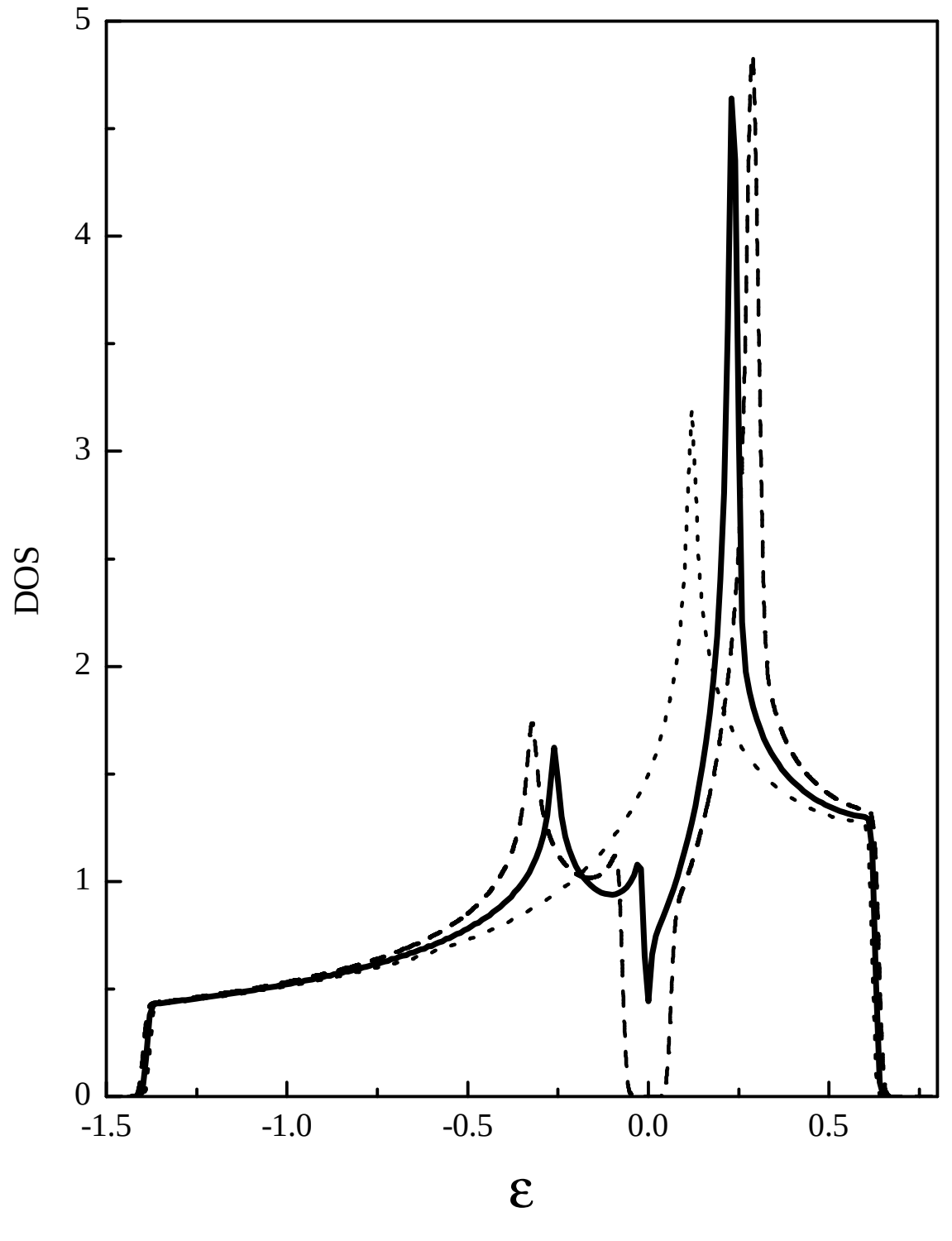

**Fig.5**